\documentclass[10pt]{article}

\usepackage[preprint]{arxiv_preprint}

\usepackage[utf8]{inputenc}
\usepackage[T1]{fontenc}
\usepackage{courier}
\usepackage{amsmath,amssymb,amsfonts}
\DeclareUnicodeCharacter{00D7}{\ensuremath{\times}}
\DeclareUnicodeCharacter{00F7}{\ensuremath{\div}}
\DeclareUnicodeCharacter{03C1}{\ensuremath{\rho}}
\DeclareUnicodeCharacter{2192}{\ensuremath{\rightarrow}}
\DeclareUnicodeCharacter{2212}{\ensuremath{-}}
\DeclareUnicodeCharacter{2264}{\ensuremath{\leq}}
\DeclareUnicodeCharacter{2265}{\ensuremath{\geq}}
\usepackage{booktabs}
\usepackage{longtable}
\usepackage{calc}
\usepackage{enumitem}
\usepackage{graphicx}
\usepackage{hyperref}
\usepackage{url}
\usepackage{microtype}
\usepackage{xcolor}
\usepackage{array}
\usepackage{tabularx}
\usepackage{tikz}
\usepackage{pgfplots}
\pgfplotsset{compat=1.18}
\usetikzlibrary{arrows.meta,calc}
\usepgfplotslibrary{groupplots}
\definecolor{NavyBlue}{RGB}{0,0,128}

\hypersetup{
  colorlinks = true,
  linkcolor  = NavyBlue,
  citecolor  = NavyBlue,
  urlcolor   = NavyBlue,
}

\providecommand{\VStd}{Standard}
\providecommand{\VDfour}{Dense-4}
\providecommand{\VDsixteen}{Dense-16}

\title{Does Tile Granularity Matter?\\
  An Empirical Study of Vector Tile Compression for Map Rendering with MapLibre Tiles}
\shorttitle{Vector Tile Compression for Map Rendering with MapLibre Tiles}
\arxivdate{October 2026 -- preprint}

\author{%
  Viacheslav Nefedov$^{1,2}$\enspace
  Ivan Nefedov$^{3}$\enspace
  Petr Datsenko$^{4}$\enspace
  Samuil Datsenko$^{4}$ \\[0.4ex]
  \normalfont\small
  $^{1}$Ariel University\quad
  $^{2}$Holon Institute of Technology\quad  
  $^{3}$Technion\quad
  $^{4}$Middle East Technical University
}


\abstract{%
Vector tiles are the display format of interactive maps: clients download the tiles that
cover the current view and render their content themselves. Their compactness matters for
storage, offline use and network transfer, particularly when maps with custom thematic
layers are served from constrained hardware or over limited links. Because vector maps
are delivered as pyramids of small, independently encoded tiles, the spatial granularity
of the encoding unit is set largely by the tiling scheme rather than by the encoding
format. We examine how this granularity affects compression in the column-oriented
MapLibre Tiles (MLT) format, using Mapbox Vector Tiles (MVT) as a feature-oriented
baseline. Five heterogeneous datasets, including a geological map and a wetland
inventory, were encoded as a conventional pyramid and as two denser variants that
represent the area of $2\times2$ and $4\times4$ conventional tiles in one independently
encoded tile at approximately the same logical level of detail. Moving from the
conventional to the $4\times4$ variant increased MLT's raw saving over MVT in all five
datasets, by 1.0--7.3 percentage points (pp). After per-tile gzip, the saving increased
in four datasets by 1.5--6.4~pp and decreased in one by 1.4~pp. Coarser granularity made
MLT streams longer in every dataset, and the gain was largest where conventional tiling
produced the shortest streams: across datasets, as the mean number of values per stream
in conventional tiles rose from 27 to 2{,}075, the raw gain fell monotonically from 7.3
to 1.0~pp. Larger units reduced structural overhead per encoded value and, most clearly
in two datasets, the encoded payload cost per value. For the two sparse datasets, coarser
units also reduced bytes per vertex for every encoding, and the $4\times4$ archives were
71\% and 33\% smaller than the conventional ones with MLT and gzip. The measured
advantage of a column-oriented vector-tile format therefore depends on the spatial
granularity at which it is evaluated, and conventional fine-grained tiling can understate
its compression potential.
}

\begin{document}
\maketitle

\section{Introduction}
\label{sec:introduction}

Interactive maps are drawn on the client from tiles. The map is cut into a pyramid of
zoom levels and, at every level, into a regular grid; each tile is encoded, stored and
transferred independently, and the client requests the tiles that cover its current view
and styles and renders their content itself \citep{osm_slippy_tiles, ogc2022tms}.
Vector tiles are therefore a display format: they carry geometry and attributes already
selected and generalized for drawing at one scale, not the source data from which they
were derived. The compactness of this representation matters well beyond download time.
It determines the storage cost of map archives that cover the whole planet,
the feasibility of offline maps on mobile devices and in low-connectivity settings,
and the volume of data that clients must transfer and decode.
These pressures grow with the underlying data. OpenStreetMap passed ten billion
nodes in August 2025 \citep{osm_wiki_stats}, and as early as 2024 the Overture Maps
release combined 2.35 billion building footprints from OpenStreetMap, Esri Community
Maps and machine-learning extractions by Microsoft and Google \citep{overture2024alpha}.

This study grew out of practical work with maps of this kind. In a commercial
fleet-management system for ground robots that operate in remote areas with little or no
connectivity, the on-site server, a computer of the Raspberry Pi class, also serves maps
to the operator. The operator views the area with custom layers that depend on the
robot's task, such as agricultural work, clearing vegetation under power lines or snow
removal. A similar need arises in a student project for the NASA International Space
Apps Challenge \citep{space_apps} on the visualization of planetary surfaces, which NASA
itself distributes as web map tiles \citep{nasa_mars_trek_wmts}. Both settings display
custom thematic layers, which are often sparse, from stored tile archives, consider MLT as
the encoding, and are constrained by storage, link capacity or computing power. Neither
system uses the results reported here, and no data from them enter the experiment. They
raised the question that this paper studies in isolation: how far the tiling scheme,
rather than the encoding format, limits how compactly such maps can be stored and
delivered.

The dominant encoding for vector tiles has been the Mapbox Vector Tile (MVT) format
\citep{mapbox_mvt_spec_2_1}, which stores each feature as a Protocol Buffers message
that holds its geometry and its attributes; in this sense MVT is feature-oriented.
MapLibre Tiles (MLT) is a recently proposed successor that reorganizes tile content
into columns: geometry and each attribute are stored in separate streams and compressed
with lightweight encodings such as dictionary, run-length and delta encoding and
integer bit-packing \citep{tremmel2025mlt, maplibre_mlt_v1}.
Column-oriented layouts with such encodings are well established in analytical
databases and data lakes \citep{abadi2006compression, zeng2023columnar, kuschewski2023btrblocks},
and spatial variants have been proposed for geospatial data lakes \citep{saeedan2022spatialparquet}.

The efficiency of columnar encoding, however, depends on how much data each
independently encoded unit contains. Every stream carries structural information, and
lightweight encodings have more opportunity to exploit regularity when a stream contains
more values. In columnar storage this dependence is well known and is controlled
explicitly: Apache Parquet lets the writer choose the size of row groups and data pages,
and its documentation notes that larger pages carry less header overhead and recommends
large row groups \citep{parquet_configurations}.

We use \emph{spatial tile granularity} in a specific sense: the geographic area and
corresponding content grouped into one independently encoded tile at an approximately
fixed logical level of detail. This is distinct from \emph{coordinate extent}, the
integer coordinate grid used inside a tile; from \emph{encoded tile size}, measured in
bytes; and from the rendered tile size in screen pixels. In a conventional pyramid the
spatial granularity is imposed by the tile matrix rather than chosen by the encoding
format. The MVT specification deliberately keeps bounds and projection outside the tile
\citep{mapbox_mvt_spec_2_1}; the quadtree pyramid, in which each zoom level doubles the
linear resolution, comes from tiling conventions shared by servers and renderers
\citep{osm_slippy_tiles, ogc2022tms}. Generalization policies keep tiles small as well,
since production tilers drop or simplify features to bound tile size.
In a measurement of seven production web maps (Appendix~\ref{app:traffic}),
the vector tiles we observed were typically only a few kilobytes on the wire,
with per-product medians between 0.1 and 3.8~KB.

Seen from the display client, this fixed unit appears smaller than necessary. Three
observations from the settings above led us to the present question. First, the rendered
tile size is fixed: MapLibre GL JS draws vector tiles at 512 CSS pixels
\citep{maplibre_gljs_vectorsource}, so at an integer zoom level a viewport of
$2560\times1600$ CSS pixels, the one used in Appendix~\ref{app:traffic}, intersects 20 to
30 tiles at once. The client thus already holds a block of several tiles for a single
view. Second, typical map tasks, such as surveying an area to choose a site for
equipment, span several screens and several zoom levels, so a session requests many more
tiles than one view: in Appendix~\ref{app:traffic}, a session of two pans and one zoom
step loaded between about 270 and 770 tiles in six of the seven products. These two
observations suggest that a coarser encoding unit need not cost much additional transfer,
although we do not measure that cost here (Section~\ref{sec:discussion}). Third, and most
important, the density of map content varies widely between tiles, whereas at a given
zoom level every tile covers the same area. Dense urban tiles are already large, while
tiles of sparse thematic layers hold a handful of features and consist largely of
per-tile structure. Because the unit is fixed by the tile matrix, an encoder cannot
enlarge it where the content is sparse. In our own map applications we have worked around
this limit with overzoom and by combining several tilesets with different maximum zoom
levels; such workarounds reduce the number of tiles but do not change the unit of
compression in a controlled way (Section~\ref{sec:bg-overzoom}).

That larger blocks of data compress better is unsurprising in itself.
Less obvious is how strongly a columnar tile format depends on spatial encoding-unit
granularity compared with a feature-oriented one, and whether the effect survives
general-purpose transport compression such as gzip \citep{rfc1951, rfc1952}. The
original MLT evaluation reports compression up to three times better than MVT across
tilesets and more than six times better for some large tiles
\citep{tremmel2025mlt}. MapLibre's current planet-scale benchmarks make the association
more explicit: naturally large tiles obtain larger MLT gains than median tiles
\citep{maplibre_mlt_benchmarks}. Those results stratify existing tiles by observed
encoded size; they do not manipulate the spatial boundary of the independently encoded
unit while holding logical level of detail approximately fixed.

Prior work has also varied vector-tile partitioning, zoom, pyramid depth and tile
content, but these interventions either change level of detail together with spatial
scale, optimize content within a fixed tile, or change pyramid organization rather than
the compression-unit boundary \citep{li2018geotile, dufilie2014feathered,
bahadori2026hifive, tippecanoe, planetiler_1574}. To our knowledge, no previous study has
systematically varied the spatial granularity of independently encoded vector tiles
under approximately matched-content, matched-LoD conditions and quantified the resulting
raw and per-tile-gzip size of column-oriented MLT relative to feature-oriented MVT.

In this paper we study this interaction empirically. We encode five heterogeneous
datasets, ranging from global small-scale reference data to large-scale national
topography, as a conventional tile pyramid and as two denser variants in which each
tile represents the content of a $2\times2$ or $4\times4$ group of conventional tiles,
with the coordinate extent increased to preserve spatial precision.
These variants are an experimental device: standard clients cannot display them
directly, but they correspond to encoding units of a quarter of a desktop view to
roughly a whole view, and they emulate such units while keeping both encoders
unchanged. For every variant we compare paired MVT and MLT tiles, as encoded
and after gzip, and relate the differences to tile content and to the internal
stream structure of MLT. We address four research questions:

\begin{enumerate}[label=\textbf{RQ\arabic*}, leftmargin=*]
  \item How does tile granularity affect the size saving of MLT over MVT,
        as encoded and after gzip?
  \item Which characteristics of tile content predict this saving?
  \item Which stream-level mechanism in MLT is consistent with the observed effect?
  \item Does coarser granularity reduce the number of bytes per vertex for both
        general-purpose (gzip) and columnar (MLT) compression?
\end{enumerate}

\noindent We make the following contributions:

\begin{enumerate}
  \item An empirical comparison of MLT and MVT across three tile granularities on five
        heterogeneous datasets and 36{,}647 tile instances, evaluated as encoded
        and after gzip, together with an openly available processing pipeline and
        per-tile statistics (Sections~\ref{sec:method} and~\ref{sec:results-granularity}).
  \item Evidence that tile content predicts raw saving strongly in four of the five
        datasets, and that across all five datasets the \VStd{} mean values per MLT
        stream ranks the raw granularity gain exactly in reverse order
        (Section~\ref{sec:results-density}).
  \item A stream-level characterization showing that coarser granularity lengthens
        MLT streams and reduces non-payload structural overhead in every dataset, while
        aggregate encoded payload bytes per value also decrease from \VStd{} to
        \VDsixteen{} in all five datasets (Section~\ref{sec:results-streams}).
  \item A format-independent comparison in bytes per vertex showing that, for sparse
        datasets, coarser granularity benefits feature-oriented, columnar and gzip
        compression alike (Section~\ref{sec:results-bpv}).
\end{enumerate}

The remainder of the paper is organized as follows.
Section~\ref{sec:background} introduces vector tiling schemes, the MVT and MLT formats
and related work. Section~\ref{sec:method} describes the datasets, the tiling variants
and the metrics. Section~\ref{sec:results} presents the results, Section~\ref{sec:discussion}
discusses their implications for tiling-scheme design, and Section~\ref{sec:threats}
examines threats to validity.

\section{Background and Related Work}
\label{sec:background}

This section introduces the concepts the study relies on: how tiled vector maps are
organized (Section~\ref{sec:bg-tiling}), how practitioners already depart from the
conventional tile pyramid (Section~\ref{sec:bg-overzoom}), how MVT and MLT lay out the
content of a tile (Section~\ref{sec:bg-encodings}), and why the size of an
independently encoded unit matters for compression (Section~\ref{sec:bg-granularity}).
Section~\ref{sec:bg-related} positions the study with respect to prior work.

\subsection{Tiled vector maps}
\label{sec:bg-tiling}

\begin{figure}[!tbp]
    \centering
    \includegraphics[width=\columnwidth]{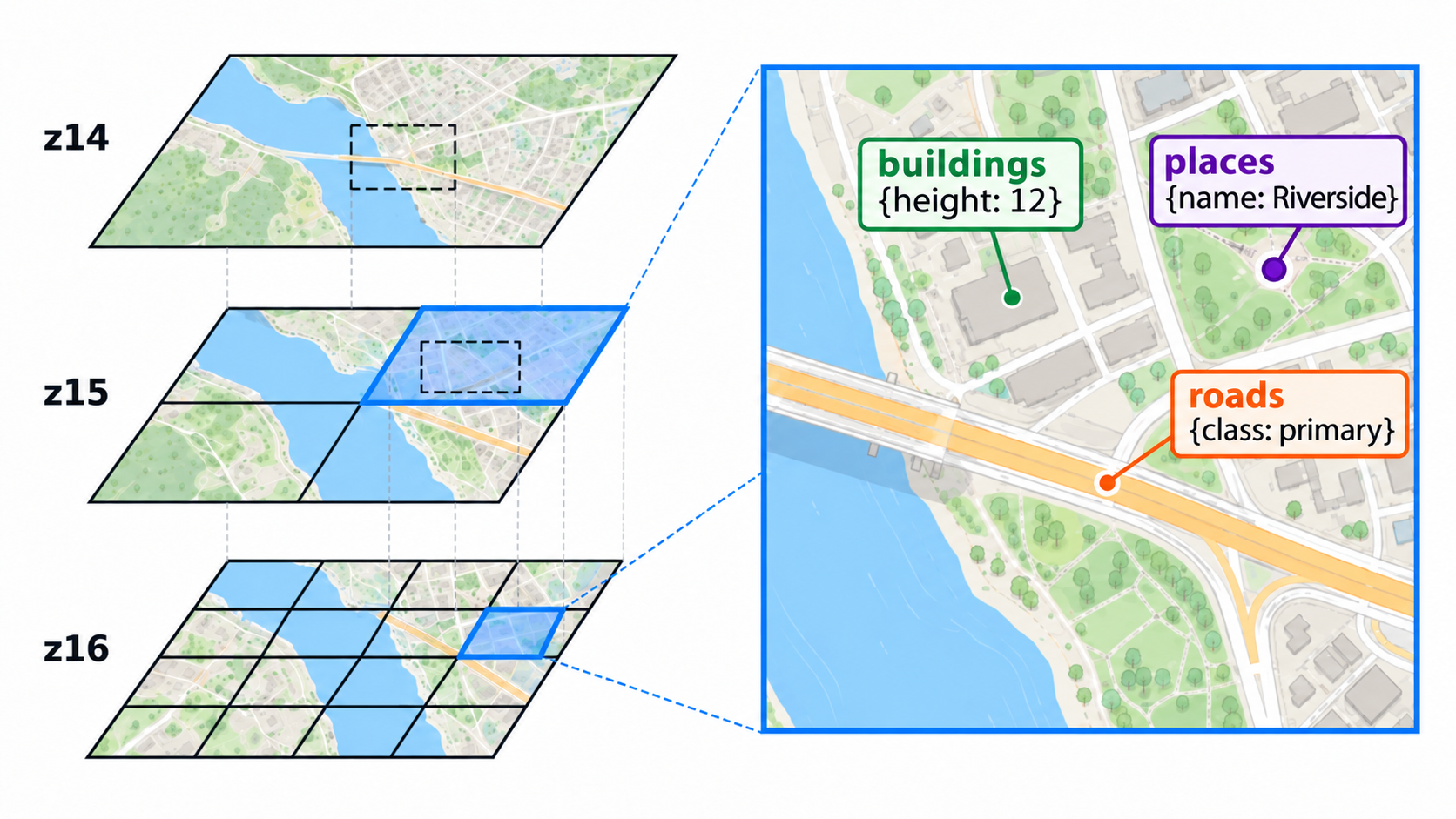}
    \caption{A tile pyramid, parent and child tiles, a viewport, and the layers,
    coordinates, features and attributes stored in one tile.}
    \label{fig:tile-concept-overview}
\end{figure}

Web maps are not delivered as a single file. The world is divided into square tiles at
a series of zoom levels, and a client downloads only the tiles that cover its current
view \citep{osm_slippy_tiles}; with vector tiles the client also styles and draws their
content itself. At zoom~0 a single tile covers the world, and each level
splits every tile into four, so a complete level~$z$ contains $4^z$ tiles. A tile is
addressed by its zoom, column and row, written $z/x/y$, and archives such as PMTiles map
each address to a stored tile \citep{pmtiles_v3}. More general tiling schemes, with other
projections and tile matrices, are standardized by the OGC Two Dimensional Tile Matrix
Set \citep{ogc2022tms}, but the quadtree scheme above is the one used by almost all web
vector maps. Figure~\ref{fig:tile-concept-overview} summarizes these concepts.

\begin{figure}[!tbp]
    \centering
    \includegraphics[width=\columnwidth]{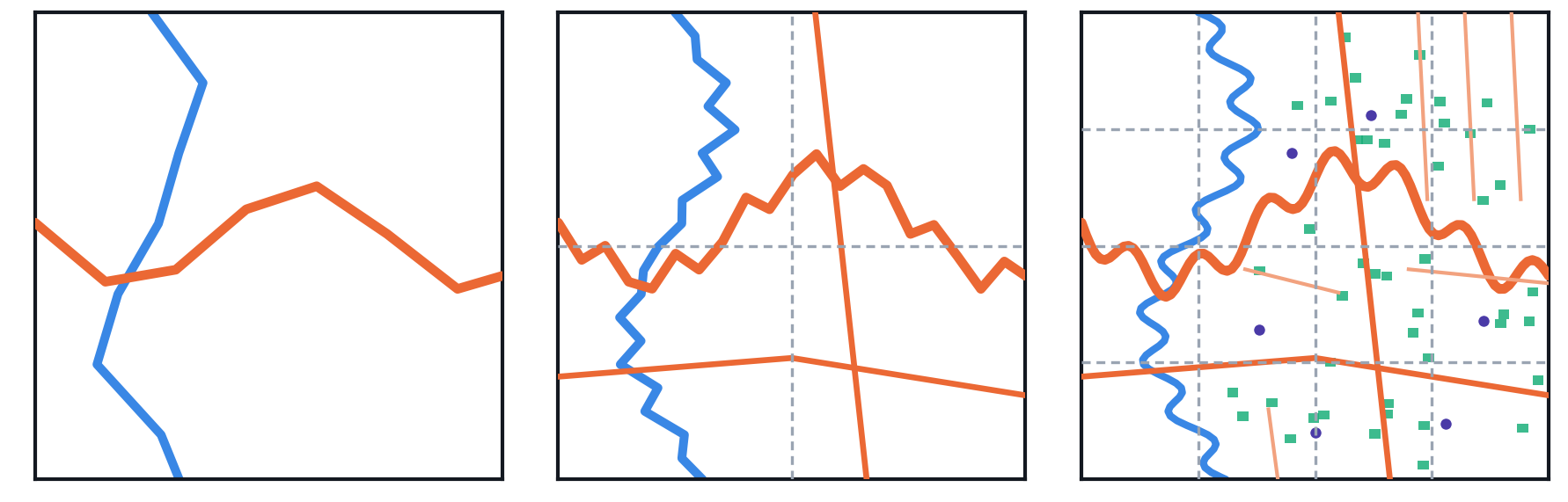}
    \caption{The same area at three zoom levels: features and detail are added as
    zoom increases.}
    \label{fig:pyramid-generalization}
\end{figure}

Each zoom level is generated for its own scale. At low zoom levels only major features
are kept and their shapes are simplified, while deeper levels add smaller features and
finer detail (Figure~\ref{fig:pyramid-generalization}). A tile is therefore a summary of
its area at one scale rather than the union of its four children. Production tile
generators also bound the size of each tile: Tippecanoe, for example, by default limits
a tile to 500~kB of compressed data and 200{,}000 features, lowers the resolution of
tiles that exceed the size limit, and offers options to drop or coalesce features
\citep{tippecanoe}.

Inside a tile, geometry is stored as integer coordinates on a local grid, the
\emph{extent}, conventionally 4096$\times$4096 units \citep{mapbox_mvt_spec_2_1}.
The tile does not record its own position: its $z/x/y$ address is kept outside the tile,
and the tile format itself does not prescribe a pyramid \citep{mapbox_mvt_spec_2_1,
maplibre_mlt_overview}. Every tile is encoded independently, so an encoder can exploit
repetition only among the features of one tile, and each tile repeats its own layer,
column or stream structure.

We distinguish four quantities that are easily conflated. \emph{Spatial tile
granularity} is the geographic area and corresponding content grouped into one
independently encoded tile at a given logical level of detail. \emph{Coordinate extent}
is the integer grid used to represent coordinates inside that tile. \emph{Encoded tile
size} is the resulting number of bytes. \emph{Rendered tile size} is the screen-space
size used by a client. Increasing coordinate extent alone changes precision, not spatial
granularity; in our experiment it is increased together with spatial aggregation so that
the coarser encoding unit retains approximately the same coordinate precision.

\subsection{Overzoom, metatiles and sparse pyramids}
\label{sec:bg-overzoom}

The dense quadtree pyramid is the default, but practitioners already depart from it,
mainly to reduce the number of stored tiles and requests.

\paragraph{Overzoom.} A client may display a tile at zoom levels deeper than the deepest
level stored, scaling the tile up instead of requesting its children. The extent is the
main limit on this practice: the extent field of MVT is an unsigned 32-bit integer whose
value 4096 is a default rather than a maximum \citep{mapbox_mvt_spec_2_1}, and MLT
declares the extent per layer and notes that a larger extent gives more precision at the
cost of more bytes \citep{maplibre_mlt_overview}. Developer proposals build on this.
A MapLibre proposal argued that ``using a large tile extent allows you to tile your
dataset only as far as is necessary'' \citep{maplibre_gljs_2507}; a Planetiler feature
request asks for an extent of 16384 at the maximum zoom so that a small tileset can be
overzoomed, instead of raising the maximum zoom, which would greatly increase the archive
size \citep{planetiler_1286}; and Tippecanoe can add extra detail at the maximum zoom to
improve precision when overzooming \citep{tippecanoe}. On the client side, MapLibre GL JS
now splits a deep-overzoomed source tile into virtual child tiles in a worker and
overscales only the last few levels, by default four \citep{maplibre_gljs_mapoptions}.

\paragraph{Metatiles.} Raster tile servers commonly render and store \emph{metatiles},
usually blocks of $8\times8$ tiles, because rendering one large area is cheaper than
rendering many small ones, labels are placed more consistently, and fewer files need to
be stored \citep{osm_wiki_metatiles}. Metatiles are a server-side unit, however: clients
still request and receive individual tiles.

\paragraph{Larger tiles.} Vector tile services have also offered larger tiles directly.
The Tilezen service (formerly Mapzen) served 256- and 512-pixel tiles, stating that
a single 512-pixel request is equivalent to four 256-pixel requests, and listed fewer
requests, better label placement, smaller files and less disk space for offline use
among the advantages of the larger size \citep{tilezen_service}. The 512-pixel vector
tile has since become the convention of WebGL renderers such as MapLibre GL JS
\citep{maplibre_gljs_vectorsource}.

\paragraph{Sparse pyramids.} Tippecanoe can generate a variable-depth pyramid that omits
child tiles wherever the parent is detailed enough to be overzoomed \citep{tippecanoe}.
A corresponding Planetiler proposal asks for variable tile depth so that descendants
need not be generated where an ancestor already represents the required detail
\citep{planetiler_1574}. Archives such as PMTiles can store such sparse pyramids, but
clients need a way to distinguish a tile that is intentionally missing, and should be
replaced by an overzoomed ancestor, from one that is empty. A design proposal for
MapLibre adds this distinction to tiled sources and TileJSON
\citep{maplibre_stylespec_1869}.

\paragraph{The rendered tile.} These approaches increase how much information a stored
tile carries, but not the size of the unit that the client renders: MapLibre GL JS, for
example, requires vector sources to use a tile size of 512 pixels
\citep{maplibre_gljs_vectorsource}. Their motivations are primarily rendering precision,
storage and request count. They do not provide a controlled measurement of compression
as a function of the spatial boundary of one independently encoded unit. The denser
variants studied here apply a related idea at every logical zoom: a tile covers the area
of multiple conventional cells while a larger coordinate extent preserves approximately
the same spatial precision.

\subsection{Feature-oriented and column-oriented tile encodings}
\label{sec:bg-encodings}

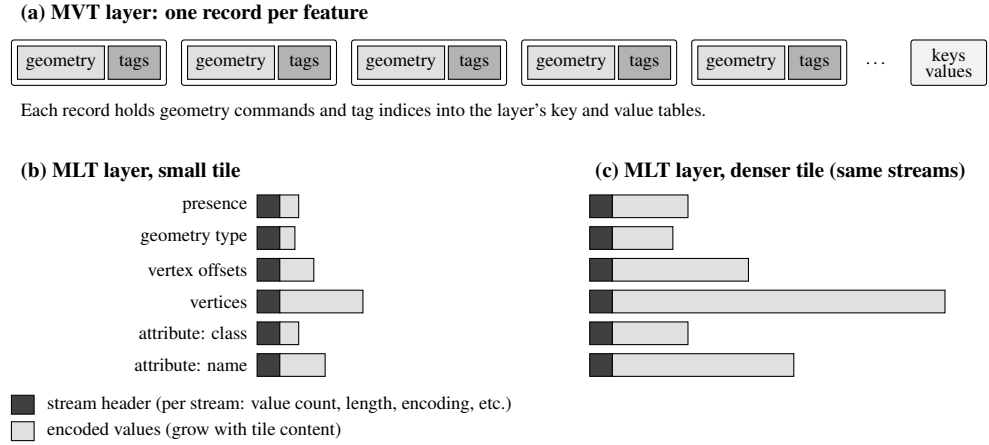
\begin{figure*}[t]
\centering
\begin{tikzpicture}[x=1cm, y=1cm, font=\scriptsize,
    hdr/.style={draw, fill=black!75, minimum height=0.32cm, inner sep=0pt},
    pay/.style={draw, fill=black!12},
    rec/.style={draw, rounded corners=1pt},
    lbl/.style={anchor=east, inner sep=2pt}]

  \node[anchor=west, font=\footnotesize\bfseries] at (-2.6,3.55)
        {(a) MVT layer: one record per feature};
  \foreach \i in {0,...,4} {
    \pgfmathsetmacro{\x}{-2.6+\i*2.25}
    \draw[rec] (\x,2.6) rectangle ++(2.05,0.6);
    \draw[pay] (\x+0.08,2.68) rectangle ++(1.15,0.44);
    \node at (\x+0.655,2.9) {geometry};
    \draw[draw, fill=black!30] (\x+1.28,2.68) rectangle ++(0.69,0.44);
    \node at (\x+1.625,2.9) {tags};
  }
  \node at (8.85,2.9) {\dots};
  \draw[rec, fill=black!5] (9.3,2.6) rectangle ++(1.0,0.6);
  \node[align=center] at (9.8,2.9) {keys\\[-2pt]values};
  \node[anchor=west] at (-2.6,2.25)
        {Each record holds geometry commands and tag indices into the layer's key and value tables.};

  \node[anchor=west, font=\footnotesize\bfseries] at (-2.6,1.45)
        {(b) MLT layer, small tile};
  \node[anchor=west, font=\footnotesize\bfseries] at (5.0,1.45)
        {(c) MLT layer, denser tile (same streams)};

  \def\names{{"presence","geometry type","vertex offsets","vertices","attribute: class","attribute: name"}}
  \def\small{{0.25,0.20,0.45,1.10,0.25,0.60}}
  \foreach \k in {0,...,5} {
    \pgfmathsetmacro{\y}{1.0-\k*0.42}
    \pgfmathparse{\names[\k]}\edef\nm{\pgfmathresult}
    \pgfmathsetmacro{\w}{\small[\k]}
    \pgfmathsetmacro{\W}{min(4*\small[\k],4.6)}
    \node[lbl] at (0.6,\y) {\nm};
    \draw[fill=black!75] (0.65,\y-0.15) rectangle ++(0.3,0.30);
    \draw[pay] (0.95,\y-0.15) rectangle ++(\w,0.30);
    \draw[fill=black!75] (5.05,\y-0.15) rectangle ++(0.3,0.30);
    \draw[pay] (5.35,\y-0.15) rectangle ++(\W,0.30);
  }

  \draw[fill=black!75] (-2.6,-1.75) rectangle ++(0.3,0.25);
  \node[anchor=west] at (-2.25,-1.625) {stream header (per stream: value count, length, encoding, etc.)};
  \draw[pay] (-2.6,-2.1) rectangle ++(0.3,0.25);
  \node[anchor=west] at (-2.25,-1.975) {encoded values (grow with tile content)};
\end{tikzpicture}
\caption{Feature-oriented and column-oriented layouts of one tile layer.
(a)~MVT stores a sequence of feature records.
(b)~MLT stores each property of all features as a column, split into streams of same-typed values;
every stream carries a header.
(c)~When a spatial encoding unit holds more content, streams become longer, amortizing per-stream
headers and exposing longer value sequences to lightweight encodings. MLT also contains layer- and
column-level structural bytes not drawn here; Section~\ref{sec:results-streams} reports total
non-payload structural share separately from stream-header share. Proportions are schematic.}
\label{fig:mvt-mlt-streams}
\end{figure*}

The Mapbox Vector Tile (MVT) format encodes a tile as a Protocol Buffers message with
named layers. Each layer holds a sequence of feature records; a record contains the
feature's geometry, written as integer drawing commands with zigzag-encoded coordinate
deltas, and its attributes, written as indices into tables of keys and values shared by
the layer \citep{mapbox_mvt_spec_2_1}. The layout is thus feature-oriented
(Figure~\ref{fig:mvt-mlt-streams}a), and redundancy across features is exploited mainly
through the shared key and value tables and, in practice, by gzip applied to the whole
tile.

MapLibre Tiles (MLT) organizes vector-tile content by column
\citep{tremmel2025mlt, maplibre_mlt_v1}. Within a layer, each property of all features,
such as geometry type, vertex offsets, vertex coordinates or an attribute, is stored as
a column, and each column is split into \emph{streams} of same-typed values, for example
presence, data, length and offset streams. Every stream carries its own metadata,
namely the number of values, the byte length and the encoding
\citep{maplibre_mlt_overview}. Streams are compressed with lightweight encodings such as
dictionary, run-length and delta encoding combined with integer packing, chosen per
stream by the encoder (Figure~\ref{fig:mvt-mlt-streams}b). The original evaluation reports
compression up to three times better than MVT over whole tilesets and more than six
times better for some large tiles \citep{tremmel2025mlt}. MapLibre's public benchmarks on
two planet-scale basemaps report smaller gains after gzip: the stable MLT~v1 reduces
gzipped archive size by about 15\% for OpenMapTiles and 5\% for Protomaps, and the
experimental MLT~v2 by about 18\% and 10\% \citep{maplibre_mlt_benchmarks}.
The same benchmarks show that disabling a dictionary shared by several columns enlarges
the uncompressed archive by about 25\% for one of the two basemaps \citep{maplibre_mlt_benchmarks},
an indication of how much the format depends on redundancy that exists within one tile.
Unlike MVT's per-feature property types, MLT declares one type per property column
within a layer of a tile; mixed input types therefore require normalization during
conversion \citep{maplibre_mlt_mvt_compat}.

\subsection{Encoding granularity in columnar storage and block compression}
\label{sec:bg-granularity}

The dependence of compression on the amount of data in one encoding unit is well known
outside cartography. Column stores compress data well because values of one column are
stored together and lightweight encodings can be applied to long runs of similar values
\citep{abadi2006compression}, and empirical studies of Parquet and ORC show that the
choice and effectiveness of encodings depend on the data within each unit
\citep{zeng2023columnar, kuschewski2023btrblocks}. These formats expose the unit size as a
parameter. Parquet lets the writer choose the size of row groups and data pages; its
documentation notes that larger pages carry less header overhead and recommends row
groups of 512~MB to 1~GB \citep{parquet_configurations}. ORC divides files into stripes of
typically about 200~MB and compresses streams in chunks of 256~kB by default, noting that
larger chunks compress better at the cost of memory \citep{orc_spec_v1}. Spatial Parquet
carries these ideas over to geospatial data lakes \citep{saeedan2022spatialparquet}.

General-purpose compressors show the same dependence. DEFLATE, the algorithm behind gzip,
finds repetitions within a sliding window of 32~kB \citep{rfc1951, rfc1952}, so data that
repeat across separately compressed files cannot be exploited. The Zstandard
documentation describes this as the small-data problem: the smaller the input, the
harder it is to compress, because at the start of a new input there is no past to build
upon, and it offers trained dictionaries, whose gains are concentrated in the first few
kilobytes, as a remedy \citep{zstd_readme}.

A vector tile differs from these settings in two ways. Its encoding unit is small,
typically kilobytes rather than megabytes (Appendix~\ref{app:traffic}), and its spatial
granularity is not a free parameter of the encoder: it follows from the tile matrix and
the generalization policy. Coordinate extent can be configured independently, but it
controls the local coordinate grid rather than the geographic area of the encoded unit.

\subsection{Related work}
\label{sec:bg-related}

Prior work approaches this question from several adjacent directions. The original MLT
evaluation compares MLT and MVT on existing tiles and reports especially large gains for
some large tiles \citep{tremmel2025mlt}. MapLibre's current planet-scale benchmarks go
further and explicitly show that naturally large tiles benefit more from MLT
\citep{maplibre_mlt_benchmarks}. These are important precedents, but tile size is an
observed property of the corpus: the spatial boundary of the independently encoded tile
is not manipulated.

Other studies manipulate quantities that are related to granularity but confound it
with another change. GeoTile reports compressed vector-tile sizes across zoom levels,
where spatial footprint changes together with feature selection and level of detail
\citep{li2018geotile}. Feathered Tiles studies adaptive spatial partitioning and targets
roughly uniform tile payloads, but does not measure a compression-versus-partition-size
curve \citep{dufilie2014feathered}. HiFIVE systematically reduces byte size by pruning
records, attributes and values within a fixed tile, so its intervention is content
reduction rather than a change of the spatial encoding unit
\citep{bahadori2026hifive}. Variable-depth pyramids and overzooming in Tippecanoe and
Planetiler change which pyramid levels are stored, not the compression boundary at a
fixed logical zoom \citep{tippecanoe, planetiler_1574, planetiler_1286}.

These distinctions matter because changing zoom, fidelity, pyramid depth or observed
payload size does not isolate spatial granularity. To our knowledge, no prior work has
systematically varied the area represented by each independently encoded vector tile
under approximately matched-content and matched-LoD conditions and quantified the
resulting raw and externally compressed size, particularly for column-oriented MLT
relative to feature-oriented MVT. Studies of columnar storage
\citep{abadi2006compression, zeng2023columnar, kuschewski2023btrblocks,
saeedan2022spatialparquet} establish the general importance of encoding-unit size, but
at much larger, non-cartographic block sizes.

\providecommand{\VStd}{Standard}
\providecommand{\VDfour}{Dense-4}
\providecommand{\VDsixteen}{Dense-16}
\providecommand{\ValidationUnchecked}{48}

\section{Data and Methods}
\label{sec:method}

The experiment converts five heterogeneous vector datasets into tile pyramids at three
granularities, encodes every pyramid in MVT and in MLT, and compares the paired
encodings tile by tile. Preparation is
split into two stages: Stage~A normalizes each raw source once into a canonical
GeoPackage, and Stage~B generates all tile variants from that GeoPackage, so that the
variants of one dataset differ only in tiling parameters.

\subsection{Datasets}
\label{sec:data}

The corpus was chosen to cover different ways in which geospatial data are produced,
because the production process largely determines how much geometry and how many
attributes a map carries per unit area. It contains a generalized small-scale reference
map (Natural Earth), a national large-scale topographic register (PDOK), two thematic
survey products (USFWS wetlands and the swisstopo geological map) and a crowdsourced
dataset (OpenStreetMap). Table~\ref{tab:datasets-provenance} lists their provenance and
Figure~\ref{fig:datasets} shows samples.

Processing whole national or global datasets at three granularities exceeded the memory
and time available for the GDAL-based pipeline, so each dataset was clipped to a study
area (Table~\ref{tab:datasets-structure}). The areas were chosen to contain dense content
for each dataset: the eastern United States for Natural Earth, the Säntis region of
north-eastern Switzerland for the swisstopo map, and the same $0.4^\circ\times0.4^\circ$
area around New Orleans for both USFWS and OSM. The last choice gives a matched pair: two
datasets with very different content over identical geography, which helps to separate
the effect of content from that of location.

\begin{table*}[!tp]
\centering
\caption{Provenance and characteristics of the five datasets.}
\label{tab:datasets-provenance}
\small
\begin{tabularx}{\textwidth}{
    @{}l
    >{\raggedright\arraybackslash}p{3.0cm}
    >{\raggedright\arraybackslash}X
    >{\raggedright\arraybackslash}p{2.1cm}
    >{\raggedright\arraybackslash}p{2.1cm}
    @{}
}
\toprule
Dataset &
Producer &
Content and production mode &
Nominal scale &
Data date \\
\midrule

Natural Earth &
NACIS &
Global physical and cultural reference layers;
generalized cartography at three scales &
1:10m--1:110m &
2021-12 \\

\addlinespace

PDOK BGT &
Dutch public authorities via PDOK &
Large-scale topographic base register
(buildings, roads, land cover, street furniture);
authoritative, continuously updated data &
1:500--1:5{,}000 &
2026-07
(snapshot, approx.) \\

\addlinespace

USFWS wetlands &
U.S.\ Fish and Wildlife Service (NWI) &
Wetland and deepwater habitat polygons for
Louisiana; Cowardin classification;
photo-interpreted thematic mapping &
1:24{,}000 &
2026-05
(release) \\

\addlinespace

swisstopo geology &
Federal Office of Topography swisstopo &
Structural Map of the Helvetic Zone,
Geological Special Map No.~128;
stratigraphic units, folds and tectonic
structures; compiled geological mapping &
1:100{,}000 &
2011
(vector edition) \\

\addlinespace

OSM Louisiana &
OpenStreetMap contributors &
Buildings, transport, land use and points
of interest for Louisiana;
crowdsourced geographic data &
Variable &
2026-09-12 \\

\bottomrule
\end{tabularx}

\vspace{2pt}
\begin{minipage}{\textwidth}
\footnotesize
\textit{Note.}
Dates refer to the dataset version or acquisition snapshot
used in the experiments, not necessarily to the dates of
original field surveys or imagery.
PDOK BGT is continuously updated; its snapshot month
is approximate.
The USFWS date denotes the May 2026 NWI release.
The swisstopo date denotes the 2011 vector edition
of Geological Special Map No.~128 (original map: 2010).
Map scales describe source cartography rather than
the zoom levels of the generated vector tiles.
\end{minipage}
\end{table*}

\begin{table*}[!tp]
\centering
\caption{Raw sources and study areas. Size, source layers, features and fields refer
to the whole raw source; fields are distinct attribute columns summed over layers (for
OSM, tag keys exposed by the GDAL OSM driver). Tile layers and the logical zoom range
refer to the tiled study area. Areas are geodesic areas of declared clipping bounds;
an em dash denotes missing bounds (PDOK). Small extent overruns for USFWS and
swisstopo are recorded in the analysis checks.}
\label{tab:datasets-structure}
\small
\setlength{\tabcolsep}{4pt}
\begin{tabular}{@{}l r r r r r r c@{}}
\toprule
 & \multicolumn{4}{c}{Raw source} & \multicolumn{3}{c}{Study area} \\
\cmidrule(lr){2-5}\cmidrule(l){6-8}
Dataset & MB & Layers & Features & Fields & km$^2$ & Tile layers & Zoom \\
\midrule
Natural Earth & 446 & 183 & 350{,}371 & 12{,}762 & 4.4\,M & 5 & 4--11 \\
PDOK BGT & 53 & 26 & 190{,}294 & 379 & --- & 10 & 3--14 \\
USFWS & 1{,}328 & 4 & 657{,}058 & 21 & 1{,}719 & 1 & 3--14 \\
swisstopo & 46 & 4 & 16{,}319 & 47 & 737 & 1 & 3--14 \\
OSM & 146 & 5 & 1{,}863{,}746 & 3{,}300 & 1{,}719 & 2 & 4--14 \\
\bottomrule

\end{tabular}
\end{table*}

\begin{table*}[!tp]
\centering
\caption{Measured structure of the normalized GeoPackages used for tile generation.
Size is the on-disk GeoPackage size in MiB; features, vertices and attribute fields
are summed over its layers.}
\label{tab:datasets-normalized}
\small
\begin{tabular}{@{}l r r r r r@{}}
\toprule
Dataset & MiB & Layers & Features & Vertices & Fields \\
\midrule
Natural Earth & 86.5 & 183 & 60{,}574 & 3{,}401{,}476 & 11{,}048 \\
PDOK BGT & 222.3 & 26 & 190{,}294 & 8{,}915{,}989 & 379 \\
USFWS & 108.6 & 4 & 41{,}781 & 5{,}840{,}438 & 21 \\
swisstopo & 0.8 & 1 & 213 & 34{,}560 & 10 \\
OSM & 14.9 & 5 & 15{,}621 & 722{,}903 & 584 \\
\bottomrule

\end{tabular}
\end{table*}

The datasets differ along the dimensions that matter for a columnar encoder.
Natural Earth has few features per unit area but many layers and attributes.
USFWS consists almost entirely of one polygon layer with very detailed boundaries,
about 210 vertices per feature in the raw source, and few attributes. PDOK is dense in
features of many classes with rich attribute schemas. The swisstopo map has few features
over a small area. OSM combines dense buildings with sparser lines and points and a very
large, sparse tag vocabulary.

Figure~\ref{fig:datasets} shows three nested views per dataset from the \VStd{}
MVT archives. The stored zoom levels and tile addresses were selected for each map's
characteristic content and fixed in a reproducible region configuration. Square
viewports lie inside both the selected tile and the dataset bounds at that zoom;
clipped margins outside those bounds are excluded. Magenta outlines mark the next
view, while zoom labels and palette explanations are typeset in LaTeX. OSM and USFWS
use the same viewports for a direct comparison of coastal Louisiana. Complete tile
addresses, viewport bounds, approximate ground sizes and color inventories are
exported with the figures.

\begin{figure*}[p]
\centering
\begingroup
\setlength{\tabcolsep}{2pt}
\renewcommand{\arraystretch}{1}
\begin{tabular}{@{}cccc@{}}
\begin{minipage}[t]{0.205\textwidth}\vspace{0pt}\centering\includegraphics[width=\linewidth]{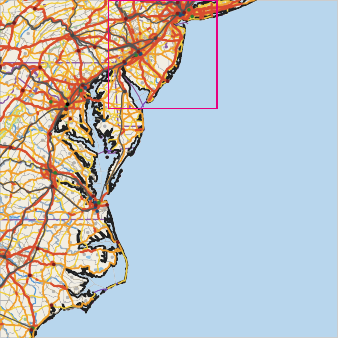}\par\vspace{1pt}\scriptsize $z=5$\end{minipage} & \begin{minipage}[t]{0.205\textwidth}\vspace{0pt}\centering\includegraphics[width=\linewidth]{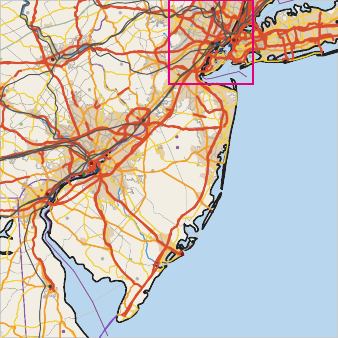}\par\vspace{1pt}\scriptsize $z=7$\end{minipage} & \begin{minipage}[t]{0.205\textwidth}\vspace{0pt}\centering\includegraphics[width=\linewidth]{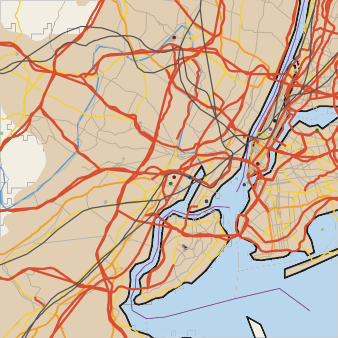}\par\vspace{1pt}\scriptsize $z=9$\end{minipage} & \begin{minipage}[t]{0.34\textwidth}\vspace{1pt}\footnotesize\raggedright \textbf{(a) Natural Earth}\\ New York / New Jersey.\par\smallskip Colors are illustrative: blue water, pale land, tan urban areas, green parks, red/orange/yellow roads, purple state boundaries and gray county boundaries. Points mark places and transport facilities. Source: Natural Earth / NACIS.\end{minipage} \\[5pt]
\begin{minipage}[t]{0.205\textwidth}\vspace{0pt}\centering\includegraphics[width=\linewidth]{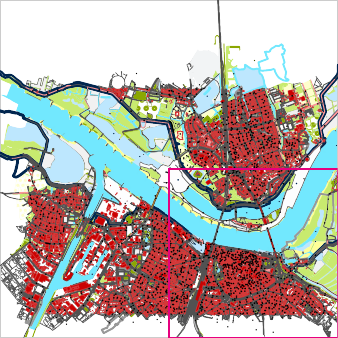}\par\vspace{1pt}\scriptsize $z=12$\end{minipage} & \begin{minipage}[t]{0.205\textwidth}\vspace{0pt}\centering\includegraphics[width=\linewidth]{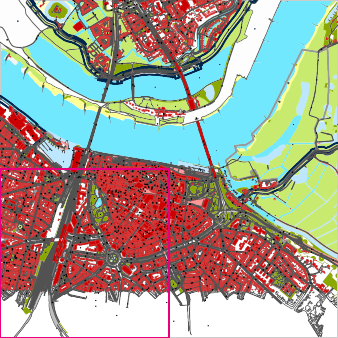}\par\vspace{1pt}\scriptsize $z=13$\end{minipage} & \begin{minipage}[t]{0.205\textwidth}\vspace{0pt}\centering\includegraphics[width=\linewidth]{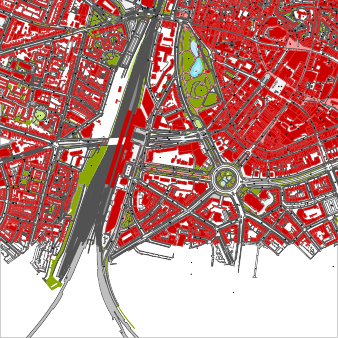}\par\vspace{1pt}\scriptsize $z=14$\end{minipage} & \begin{minipage}[t]{0.34\textwidth}\vspace{1pt}\footnotesize\raggedright \textbf{(b) PDOK BGT}\\ Nijmegen, Netherlands.\par\smallskip The palette follows the BGT standard visualization: red buildings, blue water, green vegetation and white local roads/paved areas; gray denotes classes without an assigned standard color. Source: PDOK.\end{minipage} \\[5pt]
\begin{minipage}[t]{0.205\textwidth}\vspace{0pt}\centering\includegraphics[width=\linewidth]{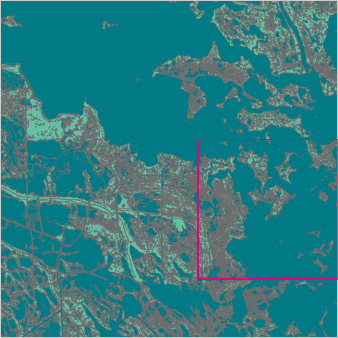}\par\vspace{1pt}\scriptsize $z=10$\end{minipage} & \begin{minipage}[t]{0.205\textwidth}\vspace{0pt}\centering\includegraphics[width=\linewidth]{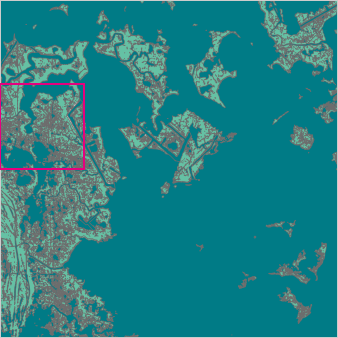}\par\vspace{1pt}\scriptsize $z=12$\end{minipage} & \begin{minipage}[t]{0.205\textwidth}\vspace{0pt}\centering\includegraphics[width=\linewidth]{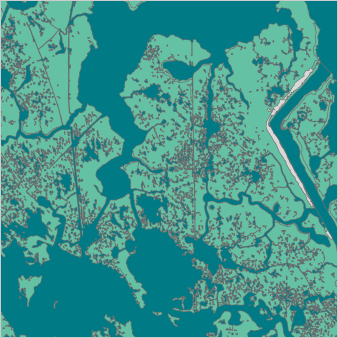}\par\vspace{1pt}\scriptsize $z=14$\end{minipage} & \begin{minipage}[t]{0.34\textwidth}\vspace{1pt}\footnotesize\raggedright \textbf{(c) USFWS wetlands}\\ coastal Louisiana.\par\smallskip The NWI palette distinguishes forested/shrub wetlands (dark green), emergent wetlands (light green), tidal wetlands (teal), and water classes (blue); unclassified polygons are light gray. Source: U.S. Fish and Wildlife Service, National Wetlands Inventory.\end{minipage} \\[5pt]
\begin{minipage}[t]{0.205\textwidth}\vspace{0pt}\centering\includegraphics[width=\linewidth]{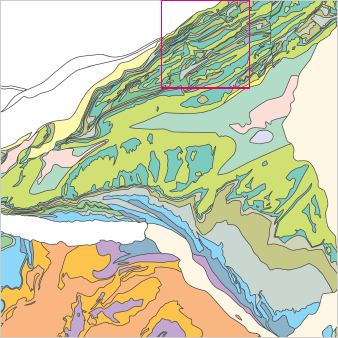}\par\vspace{1pt}\scriptsize $z=10$\end{minipage} & \begin{minipage}[t]{0.205\textwidth}\vspace{0pt}\centering\includegraphics[width=\linewidth]{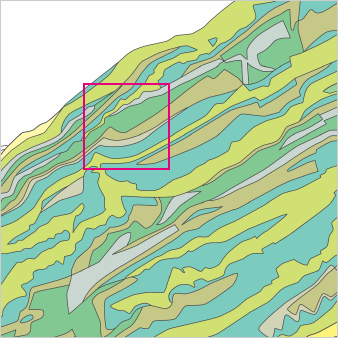}\par\vspace{1pt}\scriptsize $z=12$\end{minipage} & \begin{minipage}[t]{0.205\textwidth}\vspace{0pt}\centering\includegraphics[width=\linewidth]{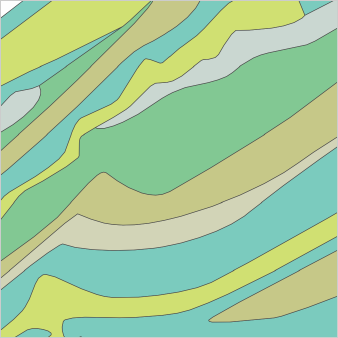}\par\vspace{1pt}\scriptsize $z=14$\end{minipage} & \begin{minipage}[t]{0.34\textwidth}\vspace{1pt}\footnotesize\raggedright \textbf{(d) swisstopo geology}\\ Saentis, Switzerland.\par\smallskip Colors distinguish geological units using the swisstopo legend and the retained \texttt{SYMBOL}/\texttt{INDEX\_} attributes. White denotes units without an assigned legend color. Source: swisstopo, Structural Map of the Helvetic Zone, sheet 128.\end{minipage} \\[5pt]
\begin{minipage}[t]{0.205\textwidth}\vspace{0pt}\centering\includegraphics[width=\linewidth]{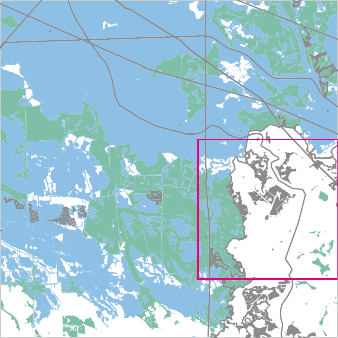}\par\vspace{1pt}\scriptsize $z=10$\end{minipage} & \begin{minipage}[t]{0.205\textwidth}\vspace{0pt}\centering\includegraphics[width=\linewidth]{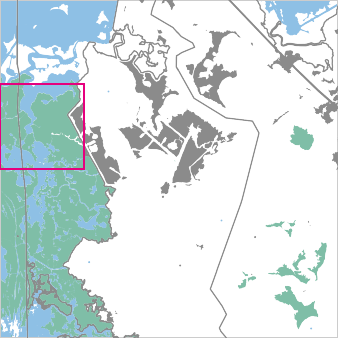}\par\vspace{1pt}\scriptsize $z=12$\end{minipage} & \begin{minipage}[t]{0.205\textwidth}\vspace{0pt}\centering\includegraphics[width=\linewidth]{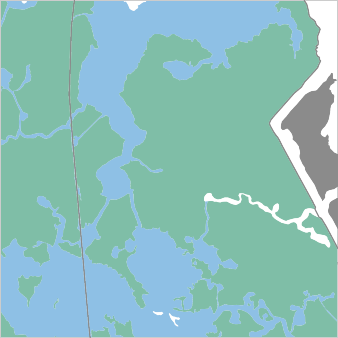}\par\vspace{1pt}\scriptsize $z=14$\end{minipage} & \begin{minipage}[t]{0.34\textwidth}\vspace{1pt}\footnotesize\raggedright \textbf{(e) OpenStreetMap}\\ coastal Louisiana.\par\smallskip An illustrative tag-based palette shows water in blue, wetlands in teal, vegetation in green, buildings in brown, roads/routes in orange and other objects in gray. Classification uses retained OSM tags rather than merged tile-layer names. Source: \copyright{} OpenStreetMap contributors, ODbL.\end{minipage} \\[5pt]
\end{tabular}
\endgroup
\caption{The five datasets at three zoom levels, rendered from Standard MVT tiles. Views are nested from left to right; magenta outlines locate the next view. Square viewports stay within the configured data bounds at each zoom. USFWS and OSM show the same coastal Louisiana areas.}
\label{fig:datasets}
\label{fig:map-natural_earth_vector}
\label{fig:map-pdok_netherlands_infrostructure}
\label{fig:map-usfws_luisiana_wetlands_geopackage}
\label{fig:map-swisstopo_mountains_vector}
\label{fig:map-louisiana-260912}
\end{figure*}

\subsection{Normalization}
\label{sec:normalization}

Each raw source was converted with GDAL into a canonical GeoPackage in Web Mercator
(EPSG:3857) with two-dimensional geometries, clipped once to the study area and indexed
spatially. Source layers were mapped onto a common set of semantic kinds (water, land,
boundary, transport, place, building, infrastructure, terrain) so that all datasets can
be rendered with one style sheet and inspected visually. Several source layers can be
merged into one tile layer, which gives between one (USFWS) and ten (PDOK) tile layers per
dataset (Table~\ref{tab:datasets-structure}). For Natural Earth, which has 183 source
layers with heterogeneous schemas, only a common subset of attributes (feature class,
names in four languages, administrative codes and rank fields) was kept. Geometry is not
altered by this mapping.
A manifest records the source and rule hashes, CRS, bounds, layer counts, feature
identity strategy and GDAL version, so that every normalized layer can be traced back to
its source.

The grouping of source layers into tile layers affects the absolute size of both
encodings, and of MLT in particular, because it determines how many features share one
set of streams. The mapping rules are identical for all variants of a dataset and for
both encodings. However, changing tile granularity can change which property types
co-occur within an output layer, so the resulting type conversions are measured
separately (Section~\ref{sec:encoding}).

\subsection{Tile generation}
\label{sec:tile-generation}

MVT tiles were generated with the GDAL MVT driver \citep{gdal_mvt} (GDAL 3.13.3, official
container image), written to MBTiles
and repacked into PMTiles archives \citep{pmtiles_v3}. We did not use Planetiler, a
common tile generator, because its output extent is fixed at 4096 units and making it
configurable is still an open request \citep{planetiler_1286}; the denser variants
require extents of 8192 and 16384. The GDAL driver exposes the extent, the buffer,
the simplification tolerance and per-layer zoom ranges as parameters. Simplification
was the same for all variants, 1.6 tile units below the maximum zoom and 1.0 at the
maximum zoom. Because the tolerance is expressed in tile units, it corresponds to the
same ground distance at the same level of detail in every variant. The buffer was
$80\cdot 2^{d}$ units, the same fraction (about 2\%) of the tile width in every variant.
The driver's limits on tile size and feature count, by default 500~kB and 200{,}000
features, were set to the largest possible value ($2^{31}-1$), so that no features were
dropped or written at reduced precision. Logical zoom ranges are listed in
Table~\ref{tab:datasets-structure}; individual tile layers start at zoom 2 to 7
depending on their content.

\subsection{Granularity variants}
\label{sec:variants}

\begin{figure}[t]
\centering
\begin{tikzpicture}[x=1cm, y=1cm, font=\scriptsize,
    unit/.style={draw=black!70, fill=black!8},
    dense/.style={draw=black, very thick, fill=black!22},
    lab/.style={align=center, text width=2.5cm, anchor=north}]
  \def\S{1.8}
  \node[anchor=west, font=\footnotesize\bfseries] at (-0.1,2.55)
    {(a) \VStd{} pyramid: every tile encoded separately};
  \foreach \n/\x/\lbl in {1/0/{zoom $z$\\1 tile}, 2/2.6/{zoom $z{+}1$\\4 tiles}, 4/5.2/{zoom $z{+}2$\\16 tiles}} {
    \pgfmathsetmacro{\c}{\S/\n}
    \foreach \i in {1,...,\n} \foreach \j in {1,...,\n}
      \draw[unit] ({\x+(\i-1)*\c},{(\j-1)*\c+0.25}) rectangle ++(\c,\c);
    \node[lab] at (\x+\S/2,0.2) {\lbl};
  }
  \node[anchor=west, font=\footnotesize\bfseries] at (-0.1,-1.25)
    {(b) One stored tile at address $z/x/y$};
  \draw[unit] (0,-3.4) rectangle ++(\S,\S);
  \node[lab] at (\S/2,-3.45) {\VStd\\content of $z$\\extent 4096};
  \draw[dense] (2.6,-3.4) rectangle ++(\S,\S);
  \draw[dashed, black!60] (2.6+\S/2,-3.4) -- ++(0,\S) (2.6,-3.4+\S/2) -- ++(\S,0);
  \node[lab] at (2.6+\S/2,-3.45) {\VDfour\\content of $z{+}1$\\extent 8192};
  \draw[dense] (5.2,-3.4) rectangle ++(\S,\S);
  \foreach \k in {1,2,3} {
    \draw[dashed, black!60] (5.2+\k*\S/4,-3.4) -- ++(0,\S);
    \draw[dashed, black!60] (5.2,-3.4+\k*\S/4) -- ++(\S,0);
  }
  \node[lab] at (5.2+\S/2,-3.45) {\VDsixteen\\content of $z{+}2$\\extent 16384};
  \foreach \x in {0,2.6,5.2}
    \draw[-{Stealth}, black!60] (\x+\S/2,-0.55) -- (\x+\S/2,-1.05);
\end{tikzpicture}
\caption{Spatial-granularity variants. (a)~In the \VStd{} pyramid an area at
logical zoom $z+1$ or $z+2$ is partitioned into 4 or 16 independently encoded tiles.
(b)~The \VDfour{} and \VDsixteen{} variants generate one independently encoded tile
over the corresponding parent area using the feature-selection and simplification rules
of logical zoom $z+1$ or $z+2$. Coordinate extent grows with the linear spatial factor
to preserve approximately the same ground-coordinate precision; extent is not itself
the granularity variable. Dashed lines mark internal \VStd{} boundaries that would
otherwise clip features and buffers.}
\label{fig:variants}
\end{figure}
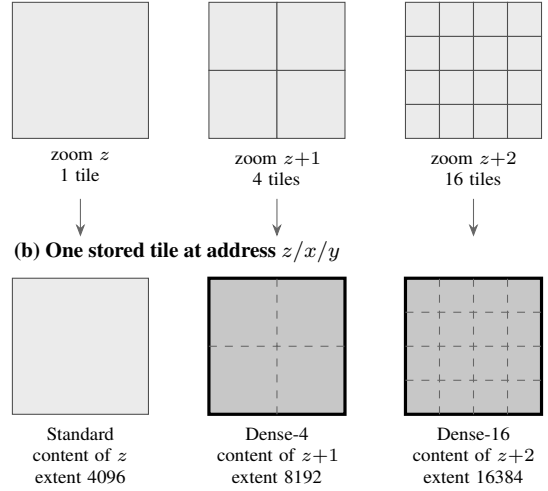

Three variants were generated from each normalized dataset
(Figure~\ref{fig:variants}, Table~\ref{tab:variants}). We operationalize \emph{spatial
tile granularity} as the area, and therefore the amount of approximately matched-LoD
content, grouped into one independently encoded tile. With detail offset
$d\in\{0,1,2\}$, one stored tile covers the area of $4^d$ \VStd{} cells at the same
logical level of detail.

In the \VStd{} variant, a tile at stored zoom $z$ is a conventional Web Mercator tile
with extent 4096. In the \VDfour{} variant, the tile stored at address $z/x/y$ is
generated using the feature selection and simplification rules of \VStd{} zoom $z+1$
over the parent area, and is encoded as one tile with extent 8192. The \VDsixteen{}
variant analogously uses the rules of \VStd{} zoom $z+2$ over the parent area and
extent 16384. Thus the coordinate resolution on the ground is approximately that of the
\VStd{} tiles at zoom $z+d$. Per-layer zoom ranges were shifted by $-d$ accordingly.
We refer to $z+d$ as the \emph{logical zoom} of a stored tile.

\begin{table}[!tbp]
\centering
\caption{Experimental spatial-granularity variants. Coordinate extent is increased with
the linear spatial factor to preserve approximately the same ground-coordinate
precision; it is not itself the granularity variable.}
\label{tab:variants}
\small
\begin{tabular}{@{}l c r r r@{}}
\toprule
Variant & $d$ & Extent & \VStd{} cells/unit & Area factor \\
\midrule
\VStd & 0 & 4{,}096 & 1 & $1\times$ \\
\VDfour & 1 & 8{,}192 & 4 & $4\times$ \\
\VDsixteen & 2 & 16{,}384 & 16 & $16\times$ \\
\bottomrule

\end{tabular}
\end{table}

The variants are an experimental manipulation of the independently encoded spatial
unit, not a proposed client rendering scheme. Spatial granularity, coordinate extent,
encoded byte size and rendered pixel size are therefore kept as separate terms
throughout the paper.

The variants preserve the intended logical level of detail, coordinate precision, and
feature-selection/simplification rules, but they do not contain exactly identical
geometry. A feature that crosses an internal boundary between the \VStd{} children of a
dense tile is stored once instead of being clipped into several pieces; moreover, the
buffer is defined relative to tile width, so changing the spatial unit changes the
ground area covered by the buffer. Feature and vertex counts therefore differ between
variants and are measured rather than forced to be equal. The primary comparison is
always MLT versus MVT \emph{within the same variant}; cross-variant archive sizes are
reported only with this caveat (Section~\ref{sec:metrics}).

\subsection{Encoding and validation}
\label{sec:encoding}

Each MLT archive is a tile-by-tile transcoding of the corresponding MVT archive.
Transcoding used the \texttt{encode\_mvt} function of the \texttt{maplibre-tiles}
Python package (version 0.2.2, installed from PyPI) with default options
\citep{maplibre_tiles_py}.

Both archives store every tile individually gzip-compressed, as is usual for tile
delivery over HTTP: MVT tiles with the default gzip settings of the GDAL PMTiles
driver and MLT tiles with zlib at level~6. Raw sizes are measured after decompression.
For the reported gzip metrics, both formats are recompressed independently at level~6
with a zero modification timestamp; every stored tile length in the corpus equals its
recompressed length, so the two measurements coincide.

\paragraph{Property types.}
MLT stores each property of a layer as one column with a single declared type,
whereas an MVT key may reference differently typed values on different features.
Mixed-type MVT properties must therefore be normalized during conversion, for example
by lossless numeric widening or by coercion to strings
\citep{maplibre_mlt_mvt_compat}. Such mixtures arise when equally named properties of
several source layers meet in one tile layer, and they become more frequent when a
coarser tile collects more features into one layer. We record the original type of
every value from the MVT protobuf and the target type of every column from the MLT
column metadata, weighted by tile-address multiplicity. Mixed-type columns occurred in
Natural Earth, PDOK and OSM, but values were coerced to strings only in two datasets:
in PDOK 1.8\% of property values in every variant, and in Natural Earth 2.5\%, 3.0\%
and 4.2\% in the \VStd{}, \VDfour{} and \VDsixteen{} variants. No values were coerced
in USFWS, swisstopo and OSM.

\paragraph{Pair audit.}
An independent audit decodes both archives of every pair and compares tile addresses,
coordinate extents, layer names, feature multiplicities, identifiers, attribute values
and geometry. All differences it found fall into two classes that follow from the
formats rather than from the transcoding: the string coercion described above, and
zero-area polygon rings, which the MVT decoder omits and the MLT decoder retains; for
the latter, the ring coordinates in the MVT command stream agree with the MLT
geometry. Apart from these two classes, the encodings agree in all 15 archive pairs.
\ValidationUnchecked{} large low-zoom Natural Earth tiles could not be decoded in MLT
by the reference decoder within the available memory and were therefore checked only
for tile addresses and sizes.

\subsection{Metrics and analysis}
\label{sec:metrics}

For a dataset, a variant $v$, and its set of tile addresses $T_v$, let
$B^{\mathrm{MVT}}_{t,v}$ and $B^{\mathrm{MLT}}_{t,v}$ be the sizes of tile $t$ in the
two encodings. The \emph{saving of MLT over MVT} is
\begin{equation}
  S(v) =
  1 - \frac{\sum_{t\in T_v} B^{\mathrm{MLT}}_{t,v}}
           {\sum_{t\in T_v} B^{\mathrm{MVT}}_{t,v}},
  \label{eq:saving}
\end{equation}
computed separately for raw tiles ($S_{\mathrm{raw}}$) and for individually gzipped
tiles ($S_{\mathrm{gz}}$). The sum form weights tiles by bytes rather than giving a tiny
and a large tile equal influence.

We call the change in this relative format advantage caused by coarser spatial
granularity the \emph{granularity gain}. For a dense variant $v$,
\begin{equation}
  G_{\mathrm{raw}}(v)
  = 100\left[S_{\mathrm{raw}}(v)-S_{\mathrm{raw}}(\mathrm{Std})\right]
  \quad\text{percentage points},
  \label{eq:granularity-gain}
\end{equation}
and $G_{\mathrm{gz}}(v)$ is defined analogously after gzip. Thus the headline
``raw granularity gain'' is $G_{\mathrm{raw}}(\mathrm{D16})$. It is a
difference of two \emph{within-variant} MLT-versus-MVT savings, not the percentage
reduction in total archive size between \VStd{} and \VDsixteen{}. This distinction is
important because clipping and buffering make the cross-variant feature and vertex
counts slightly different.

Tiles that PMTiles stores once for several addresses are counted at every address in
the tile-size metrics, matching a request for each address. Raw sizes are measured after
tile decompression; gzip sizes are the individually compressed tile sizes.

Tile content is described by the number of decoded features and vertices. At tile level
we use
$s_{t,v}=1-B^{\mathrm{MLT}}_{t,v}/B^{\mathrm{MVT}}_{t,v}$ and relate it to feature and
vertex count using logarithmic bins and Spearman rank correlations computed within each
dataset across all individual tile addresses from all three variants. To compare
variants at the same intended level of detail, we also aggregate savings by
\emph{logical zoom} $z+d$.

For the stream-level analysis we distinguish payload from structure. For each parsed MLT
layer frame, $P$ is the sum of encoded stream payload bytes, $H_s$ is the sum of stream
header bytes, and $F$ is the complete layer-frame size. We define
\begin{equation}
  H_{\mathrm{struct}} = F-P,
  \qquad
  R_{\mathrm{struct}} = \frac{F-P}{F},
  \label{eq:structural-share}
\end{equation}
where $H_{\mathrm{struct}}$ contains layer, column and stream framing, and $H_s$ is a
subset of it. We also report mean values per stream,
$\sum n_{\mathrm{values}}/N_{\mathrm{streams}}$, and aggregate encoded payload bytes per
value, $\sum P/\sum n_{\mathrm{values}}$. The latter is a descriptive measure of the
bytes used by encoded stream payloads, not a codec-normalized entropy estimate.

PMTiles can map several addresses to one stored payload. Structural stream statistics
are therefore collected once per stored payload, whereas the headline $S$ and $G$
metrics are address-weighted. We use the stream statistics to characterize mechanisms
and do not algebraically decompose $G_{\mathrm{raw}}$ into header and payload terms.
Finally, to compare all four byte representations independently of MLT/MVT pairing, we
compute \emph{bytes per stored vertex}, $\sum_t B_t/\sum_t V_t$, for raw MVT, gzipped
MVT, raw MLT and gzipped MLT in every variant.

The processing pipeline, the normalization manifests and the per-tile and per-stream
statistics are openly available (see Data Availability).
%

\section{Results}
\label{sec:results}
\providecommand{\PaperTilesTotal}{36{,}647}
\providecommand{\PaperNaturalEarthTileShare}{93}

\subsection{Effect of granularity}
\label{sec:results-granularity}

Table~\ref{tab:main} summarizes the saving of MLT over MVT for every dataset and
variant. The raw granularity gain
$G_{\mathrm{raw}}(\mathrm{D16})=S_{\mathrm{raw}}(\mathrm{D16})-
S_{\mathrm{raw}}(\mathrm{Std})$ was positive in all five datasets, from 1.0 to 7.3
percentage points (mean 4.3~pp). After gzip, $G_{\mathrm{gz}}(\mathrm{D16})$ was
positive in four datasets, from 1.5 to 6.4~pp, and negative for swisstopo
($-1.4$~pp; mean over all five datasets 2.6~pp). In every dataset the \VDfour{}
variant lies between the other two for the raw saving.

\begin{table*}[!tp]
\centering
\caption{Saving of MLT over MVT (\%, Equation~\ref{eq:saving}) by dataset and variant,
for raw tiles and after gzip. $G$ is the granularity gain from \VStd{} to
\VDsixteen{} (Equation~\ref{eq:granularity-gain}), in percentage points. Tiles are
counted per address.}
\label{tab:main}
\small
\begin{tabular}{@{}l r r r r r r r r r r r@{}}
\toprule
 & \multicolumn{3}{c}{Tiles} & \multicolumn{4}{c}{$S_{\mathrm{raw}}$} & \multicolumn{4}{c}{$S_{\mathrm{gz}}$} \\
\cmidrule(lr){2-4}\cmidrule(lr){5-8}\cmidrule(l){9-12}
Dataset & Std & D4 & D16 & Std & D4 & D16 & $G_{\rm raw}$ & Std & D4 & D16 & $G_{\rm gz}$ \\
\midrule
Natural Earth & 25{,}800 & 6{,}554 & 1{,}705 & 38.6 & 41.1 & 45.9 & $+7.3$ & 8.6 & 11.4 & 15.0 & $+6.4$ \\
PDOK BGT & 153 & 70 & 39 & 61.7 & 62.1 & 62.7 & $+1.0$ & 17.5 & 18.8 & 20.3 & $+2.8$ \\
USFWS & 606 & 190 & 81 & 44.4 & 45.8 & 46.9 & $+2.4$ & 13.8 & 14.5 & 15.2 & $+1.5$ \\
swisstopo & 435 & 130 & 50 & 19.3 & 24.5 & 25.8 & $+6.5$ & 11.9 & 12.7 & 10.5 & $-1.4$ \\
OSM & 576 & 182 & 76 & 46.5 & 49.1 & 50.6 & $+4.1$ & 9.9 & 11.8 & 13.6 & $+3.8$ \\
\bottomrule

\end{tabular}
\end{table*}

\begin{figure*}[!tp]
\centering
\includegraphics[width=\textwidth]{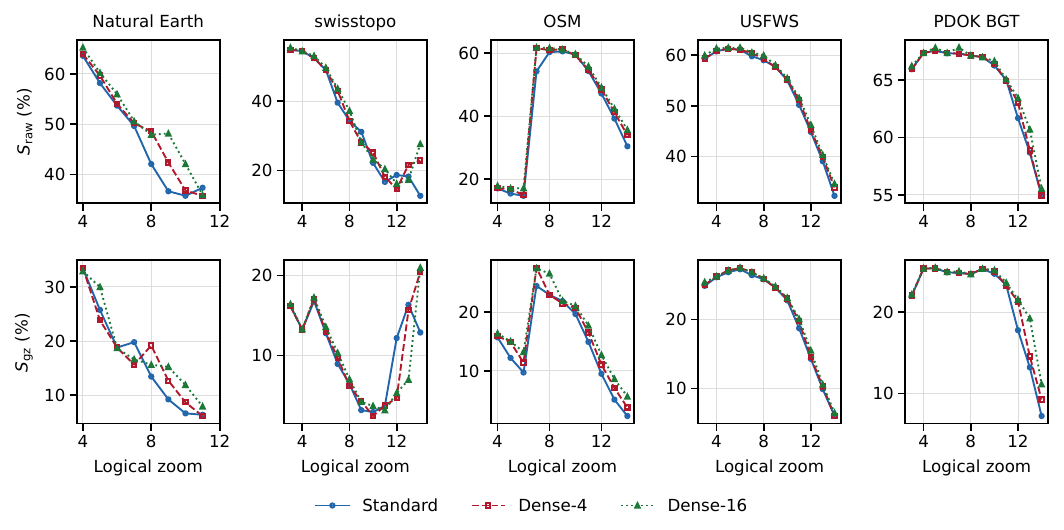}
\caption{Saving of MLT over MVT by logical zoom, raw (top) and after deterministic gzip-6 (bottom), for the
three variants. Datasets are ordered from sparse to dense conventional tiles (values per
MLT stream). At the same logical zoom the variants use the same intended level-of-detail rules; the dense
variants cover the area of 4 or 16 conventional cells per tile.}
\label{fig:zoom-aligned}
\end{figure*}

The archive-level numbers are dominated by the zoom levels that hold most bytes.
Figure~\ref{fig:zoom-aligned} therefore compares the variants at the same logical zoom,
where they use the same intended feature-selection and simplification regime while one
independently encoded tile covers 1, 4 or 16 conventional cell areas. Across the 55 combinations of dataset and logical zoom, the \VDsixteen{} saving
was higher than the \VStd{} saving in 46 cases and lower in 4 for raw tiles, and higher in
44 and lower in 6 after gzip; for OSM and USFWS it was higher at every zoom level; in the remaining cases the difference was below 0.05~pp, typically at low zoom
levels where the whole study area fits into a single tile in every variant.
The gains are largest at the deeper logical zoom levels, where the \VStd{} tiles are
smallest; for Natural Earth at logical zoom 9, for example, the raw saving rises from
36.7\% to 48.2\%.

\subsection{Tile content and the size of the effect}
\label{sec:results-density}

Pooling all tiles, the saving grows with the content of the tile
(Table~\ref{tab:bins}). Tiles with fewer than 100 vertices were on average larger in MLT
than in MVT, by 6.2\% raw and 4.1\% after gzip (mainly Natural Earth, $-6.5\%$, and
USFWS, $-10.2\%$; in the other three datasets such tiles saved less than 5\%), while tiles with more than 100{,}000
vertices were 58.6\% smaller raw and 18.3\% after gzip. The rank correlation between the
mean vertex count of a bin and its saving is 0.93 raw and 0.98 after gzip; for feature
count it is 1.0 for both.

\begin{table}[!tbp]
\centering
\caption{Saving of MLT over MVT by vertices per tile, pooled over all tiles of the five
datasets and three variants (\PaperTilesTotal{} tiles,
\PaperNaturalEarthTileShare\% of them from Natural Earth).}
\label{tab:bins}
\small
\setlength{\tabcolsep}{4pt}
\begin{tabular}{@{}l r r r r@{}}
\toprule
Vertices per tile & Tiles & Mean MVT (KiB) & $S_{\mathrm{raw}}$ & $S_{\mathrm{gz}}$ \\
\midrule
1--99 & 5{,}902 & 0.7 & $-6.2$ & $-4.1$ \\
100--499 & 15{,}584 & 19.0 & $37.8$ & $6.6$ \\
500--999 & 7{,}331 & 23.2 & $35.7$ & $7.0$ \\
1{,}000--4{,}999 & 5{,}845 & 32.3 & $36.3$ & $8.7$ \\
5{,}000--9{,}999 & 830 & 62.0 & $41.8$ & $11.3$ \\
10{,}000--49{,}999 & 779 & 151.1 & $50.2$ & $15.9$ \\
50{,}000--99{,}999 & 173 & 459.6 & $51.6$ & $15.6$ \\
$\geq$100{,}000 & 203 & 3{,}073.6 & $58.6$ & $18.3$ \\
\bottomrule

\end{tabular}
\end{table}

Because 93\% of the pooled tiles come from Natural Earth, we also computed the relation
within each dataset over its individual tile addresses in all three variants. For raw
tiles, per-tile saving $s_t$ rises strongly with feature count in OSM, PDOK, swisstopo
and USFWS (Spearman $\rho=0.92$--$0.96$) and with vertex count
($\rho=0.68$--$0.91$). Natural Earth is the exception ($\rho=0.28$ for features and
0.27 for vertices): tiles with similar counts can have very different layer and
attribute composition. After gzip, the feature-count correlations are
0.52--0.65 and the vertex-count correlations 0.10--0.66. They are weaker than the raw
correlations in the four datasets above, but stronger than Natural Earth's weak raw
relation. Tile content is therefore a strong raw predictor within four datasets, but
count alone is not a universal predictor across schemas or layer mixtures.

Across datasets, a quantity closer to the columnar encoder gives a much cleaner ordering:
the mean number of values per MLT stream in the \VStd{} variant is 27 for Natural Earth,
44 for swisstopo, 341 for OSM, 980 for USFWS and 2{,}075 for PDOK. Their raw
granularity gains are respectively 7.3, 6.5, 4.1, 2.4 and 1.0~pp---the exact reverse
rank ($\rho=-1.0$ across the five datasets). The \VStd{} share of parsed MLT layer-frame
bytes outside encoded stream payloads gives the same ordering in the opposite direction:
44.0\%, 10.0\%, 2.50\%, 0.99\% and 0.46\%, respectively
(Table~\ref{tab:streams}). With only five heterogeneous datasets this rank relation is
descriptive rather than an inferential model, but it identifies fragmentation of the
columnar representation as a stronger cross-dataset indicator of raw granularity gain
than feature or vertex count alone.

The same mechanism is visible within Natural Earth. In the \VStd{} variant the sparse
layers \emph{places}, \emph{hydrography} and \emph{land use} were 3.5\%, 3.3\% larger
and 0.1\% smaller in MLT than in MVT, whereas in the \VDsixteen{} variant they were
35.8\%, 16.4\% and 16.5\% smaller.

\subsection{Stream-level mechanism}
\label{sec:results-streams}

The structural parser successfully decoded every MLT layer frame in all five datasets
and variants. Table~\ref{tab:streams} reports three complementary quantities. First,
\emph{values per stream} measures how fragmented the columnar representation is.
Second, the \emph{structural share} $R_{\mathrm{struct}}$ is the fraction of parsed
layer-frame bytes outside encoded stream payloads; it includes layer, column and stream
framing. Third, payload bytes/value measures the aggregate size of the encoded stream
payloads themselves. These quantities are descriptive structural measurements; because
PMTiles may map several addresses to one stored payload, we do not use them as an exact
algebraic decomposition of the address-weighted granularity gain.

Coarser granularity lengthens streams in every dataset. From \VStd{} to
\VDsixteen{}, mean values per stream rise from 27 to 89 for Natural Earth, 44 to 338 for
swisstopo, 341 to 1{,}743 for OSM, 980 to 7{,}135 for USFWS, and 2{,}075 to
5{,}586 for PDOK. The structural share falls monotonically in all five: from 44.0\% to
19.6\%, 10.0\% to 2.27\%, 2.50\% to 0.61\%, 0.99\% to 0.14\%, and 0.46\% to
0.17\%, respectively. The subset consisting only of stream headers falls as well
(for example, from 21.0\% to 9.5\% in Natural Earth). Thus one directly measured
mechanism is amortization of MLT structure over longer value sequences.

The encoded payloads show a second effect. Aggregate payload bytes/value decline from
\VStd{} to \VDsixteen{} in all five datasets. The change is large for swisstopo
(1.973 to 1.287 bytes/value, $-35\%$) and OSM (0.922 to 0.802, $-13\%$), but small
for Natural Earth (0.435 to 0.429), USFWS (0.895 to 0.879) and PDOK (1.342 to
1.315). Natural Earth's large $G_{\mathrm{raw}}$ therefore coincides with a very large
reduction in structural share and almost unchanged payload bytes/value, whereas
swisstopo and OSM show substantial changes in the encoded payload as well. Conversely,
PDOK and USFWS already have long streams and sub-1\% structural shares in \VStd{}, and
their raw granularity gains are the two smallest. This is the stream-level evidence
behind RQ3: larger spatial units both amortize structural overhead and can improve the
encoding of longer value sequences, but the balance between the two mechanisms is
dataset-dependent.

\begin{table*}[!tp]
\centering
\caption{MLT stream structure by granularity. Structural share is
$(F-P)/F$ (Equation~\ref{eq:structural-share}), where $F$ is parsed layer-frame bytes
and $P$ encoded stream payload bytes. Payload B/value is $\sum P/\sum n_{\rm values}$.
$G_{\rm raw}$ is the raw granularity gain from \VStd{} to \VDsixteen{}.}
\label{tab:streams}
\footnotesize
\setlength{\tabcolsep}{3.2pt}
\begin{tabular}{@{}l r r r r r r r r r@{}}
\toprule
& \multicolumn{3}{c}{Values/stream}
& \multicolumn{3}{c}{Structural share (\%)}
& \multicolumn{2}{c}{Payload B/value}
& $G_{\rm raw}$ \\
\cmidrule(lr){2-4}\cmidrule(lr){5-7}\cmidrule(lr){8-9}
Dataset & Std & D4 & D16 & Std & D4 & D16 & Std & D16 & (pp) \\
\midrule
Natural Earth & 27 & 41 & 89 & 44.03 & 33.89 & 19.56 & 0.435 & 0.429 & 7.3 \\
swisstopo & 44 & 133 & 338 & 10.03 & 4.77 & 2.27 & 1.973 & 1.287 & 6.5 \\
OSM & 341 & 864 & 1{,}743 & 2.50 & 1.13 & 0.61 & 0.922 & 0.802 & 4.1 \\
USFWS & 980 & 2{,}992 & 7{,}135 & 0.99 & 0.34 & 0.14 & 0.895 & 0.879 & 2.4 \\
PDOK BGT & 2{,}075 & 3{,}896 & 5{,}586 & 0.46 & 0.25 & 0.17 & 1.342 & 1.315 & 1.0 \\
\bottomrule

\end{tabular}
\end{table*}

\subsection{Bytes per vertex}
\label{sec:results-bpv}

Table~\ref{tab:bpv} compares the four encodings independently of the pairing. For the
two sparse datasets, coarser granularity reduced the bytes per vertex of every encoding:
for Natural Earth by 67\% (MVT raw) to 60\% (MLT with gzip) between \VStd{} and
\VDsixteen{}, and for swisstopo by 33\% to 29\%. For the three datasets with dense
tiles the bytes per vertex changed by less than 10\% (for OSM, MLT fell by 9\% raw and
MVT by 2\%), and MLT kept its advantage.
Part of the reduction for Natural Earth comes from features that are no longer clipped
into many tiles: the number of stored features fell from 2.0 to 0.7 million and the
number of stored vertices by 28\%, so each remaining vertex carries less repeated
attribute data.

In absolute terms, the \VDsixteen{} archive stores the same logical zoom range as the
\VStd{} archive. For the two sparse datasets it was much smaller: with MLT and gzip,
71\% smaller for Natural Earth (41.0 against 142.8~MiB) and 33\% smaller for swisstopo;
with MVT and gzip, 69\% and 34\% smaller. For the three datasets with dense tiles the
size changed little: with MLT and gzip $-5\%$ for OSM and PDOK and $+2\%$ for USFWS. Because the dense variants
do not repeat features clipped at internal tile boundaries, these totals include the
effect of less duplication as well as of better compression; both are consequences of
coarser tiles.

\begin{table}[!tbp]
\centering
\caption{Bytes per stored vertex for MVT and MLT, raw and gzipped.}
\label{tab:bpv}
\footnotesize
\setlength{\tabcolsep}{3.5pt}
\begin{tabular}{@{}l l r r r r@{}}
\toprule
Dataset & Variant & MVT & MVT+gz & MLT & MLT+gz \\
\midrule
Natural Earth & \VStd & 29.8 & 9.1 & 18.3 & 8.3 \\
 & \VDfour & 15.4 & 5.4 & 9.1 & 4.8 \\
 & \VDsixteen & 9.9 & 3.9 & 5.3 & 3.3 \\
\addlinespace
PDOK BGT & \VStd & 15.2 & 6.2 & 5.8 & 5.1 \\
 & \VDfour & 15.1 & 6.2 & 5.7 & 5.1 \\
 & \VDsixteen & 15.3 & 6.3 & 5.7 & 5.0 \\
\addlinespace
USFWS & \VStd & 4.3 & 2.6 & 2.4 & 2.2 \\
 & \VDfour & 4.3 & 2.6 & 2.3 & 2.2 \\
 & \VDsixteen & 4.4 & 2.6 & 2.3 & 2.2 \\
\addlinespace
swisstopo & \VStd & 5.4 & 3.9 & 4.4 & 3.4 \\
 & \VDfour & 4.1 & 3.0 & 3.1 & 2.6 \\
 & \VDsixteen & 3.6 & 2.7 & 2.7 & 2.4 \\
\addlinespace
OSM & \VStd & 6.1 & 2.8 & 3.2 & 2.5 \\
 & \VDfour & 5.9 & 2.6 & 3.0 & 2.3 \\
 & \VDsixteen & 5.9 & 2.6 & 2.9 & 2.3 \\
\bottomrule

\end{tabular}
\end{table}

\subsection{The after-gzip exception}
\label{sec:results-gzip}

The only negative after-gzip gain, for swisstopo, comes from a small archive: 435
\VStd{} tiles with 0.64~MiB of gzipped MVT in total. At matched logical zoom the
\VDsixteen{} variant is better than \VStd{} at eight zoom levels and worse at three,
most strongly at logical zooms 12 and 13 (5.3\% and 7.0\% against 12.2\% and 16.4\%),
whereas at zoom 14, which holds most tiles, it is better (21.0\% against 12.9\%).
With so few bytes, the archive-level difference of 1.4~pp corresponds to a few
kilobytes, and we do not interpret it further than as evidence that the after-gzip
benefit is not guaranteed.


\section{Discussion}
\label{sec:discussion}

\paragraph{Why granularity matters for a columnar format.}
The stream measurements support two mechanisms rather than a single metadata-versus-data
split. First, coarser spatial units amortize MLT structure. The share of parsed layer
frames outside encoded stream payloads falls in every dataset, from 44.0\% to 19.6\%
for Natural Earth and from 10.0\% to 2.27\% for swisstopo, while it is already below
1\% in \VStd{} USFWS and PDOK. Second, longer streams can be encoded more compactly:
aggregate payload bytes/value fall from \VStd{} to \VDsixteen{} in all five datasets,
most strongly for swisstopo and OSM. Natural Earth is the clearest structural-overhead
case: it gains 7.3~pp raw while payload bytes/value is nearly unchanged. These are
descriptive mechanisms, not an exact algebraic decomposition of $G_{\mathrm{raw}}$,
because stream statistics are collected once per stored payload whereas the headline
size metric is counted per tile address (RQ1, RQ3).
A third, smaller factor applies only to Natural Earth, the one dataset in which string
coercion of mixed-type properties grows with granularity, from 2.5\% to 4.2\% of values
(Section~\ref{sec:encoding}); its gain therefore also includes the encoder's response to
these mixtures. In the other four datasets the coercion share is zero or constant across
variants.

\paragraph{What predicts the raw granularity gain.}
Feature and vertex count are useful within four datasets, but they are not
schema-independent predictors. The cleanest cross-dataset ordering is the fragmentation
of the MLT representation itself. \VStd{} mean values per stream increase from 27
(Natural Earth) to 44 (swisstopo), 341 (OSM), 980 (USFWS) and 2{,}075 (PDOK), while
$G_{\mathrm{raw}}$ decreases in exactly that order from 7.3 to 6.5, 4.1, 2.4 and
1.0~pp. The \VStd{} structural share gives the same rank in the other direction.
With only five deliberately heterogeneous datasets this should not be read as a fitted
predictive law, but it supports a practical rule: spatial aggregation has the most room
to help MLT when the conventional tiling scheme fragments columns into many short
streams with substantial structure per payload byte (RQ2).

\paragraph{After gzip.}
Gzip removes much of the redundancy that MLT exploits, so the after-gzip savings are
smaller than the raw savings in every dataset, between 8.6\% and 17.5\% for the \VStd{}
variant. This range is consistent with the 5--18\% reductions that MapLibre reports for
gzipped planet-scale archives \citep{maplibre_mlt_benchmarks}, which suggests that our
study areas are not atypical. The granularity effect survives gzip in four of five
datasets and in most matched zoom levels, but it is smaller and not guaranteed.

\paragraph{Implications for tiling schemes.}
The results suggest that the tiling scheme should be chosen together with the encoding.
Settings in which retrieval granularity matters less than size, such as offline
archives, bulk distribution or server-side caches, can directly consider coarser
independently encoded units. For interactive delivery, overzoom, larger coordinate
extents and variable-depth or sparse pyramids are relevant engineering mechanisms
(Section~\ref{sec:bg-overzoom}), but they are not equivalent to our controlled
aggregation: omitting descendant tiles changes pyramid organization, whereas our
experiment changes the compression-unit boundary at a matched logical zoom. Whether
those production mechanisms reproduce the measured compression gain is therefore a
separate question. The vector tiles served by the production web maps we observed were
small (Appendix~\ref{app:traffic}), which makes this interaction practically relevant.

\paragraph{The cost of coarser tiles.}
A coarser tile covers four or sixteen times the area of a conventional one. A client that
needs a small region at a given zoom therefore downloads content outside its viewport,
and a change of view invalidates more data at once. In the viewport of
Appendix~\ref{app:traffic}, a \VDfour{} unit drawn at its logical zoom spans
$1024\times1024$ CSS pixels and fits within one view, whereas a \VDsixteen{} unit spans
$2048\times2048$ CSS pixels, more than the height of the viewport; its overfetch can be
offset only over sessions that cover several views. Whether the smaller total size
outweighs this overfetch depends on the viewport, the zoom behaviour of users and caching,
and we did not measure it. Our variants therefore show what granularity can offer in
compression, not a recommendation for a single tile size: the favourable range depends
on the density of the content and on how much of the area a session covers.

\paragraph{Beyond MLT.}
For the two sparse datasets, bytes per vertex fell for all four encodings, including
gzipped MVT (Table~\ref{tab:bpv}). The dependence of compression on the encoding unit is
therefore not specific to MLT; MLT is more sensitive to it because of its per-stream
structure. Any future encoding that relies on redundancy within a tile will face the
same constraint, and its evaluation should state the tiling under which it was measured
(RQ4).

\paragraph{Back to the motivating settings.}
The settings that motivated this study (Section~\ref{sec:introduction}) display sparse
custom thematic layers from stored tile archives, and they correspond to the favourable
end of our results. The two sparse datasets, Natural Earth and the swisstopo geological
map, had the largest raw granularity gains (7.3 and 6.5~pp), and their \VDsixteen{}
archives were 71\% and 33\% smaller than the \VStd{} archives with MLT and gzip
(Section~\ref{sec:results-bpv}). At \VStd{} granularity the sparse Natural Earth layers
\emph{places} and \emph{hydrography} were even larger in MLT than in MVT, so the fixed
unit can not only limit the advantage of the columnar format but reverse it
(Section~\ref{sec:results-density}). A geological map is also the kind of thematic
layer used for planetary surfaces. The after-gzip picture is less clear: for swisstopo
the advantage of MLT over MVT after gzip decreased slightly with coarser units
(Section~\ref{sec:results-gzip}), so most of its archive reduction came from coarser
tiles as such rather than from the columnar format. These results do not describe the
behaviour of such systems, since we measured neither serving cost on low-power
hardware, nor decoding cost on the client, nor overfetch. They indicate that for maps of
this kind the tiling scheme, and not only the encoding format, should be revisited when
compactness matters.

\section{Threats to Validity}
\label{sec:threats}

\paragraph{Construct validity.}
The dense variants emulate larger encoding units; they are not identical in content to
the \VStd{} pyramid. Features crossing internal tile boundaries are not clipped, and the
buffer covers a different ground area, so the numbers of stored features and vertices
differ between variants (for Natural Earth the stored features fall by two thirds). We
therefore base the main comparison on the two encodings within each variant, add bytes
per vertex, and report total archive sizes only with this caveat. Grouping source layers
into tile layers and, for Natural Earth, reducing the attribute set were needed for a
common style; they affect absolute savings. The mapping rules are identical across
variants and encodings, but the property types that meet in one tile layer can change
with granularity; MLT's single type per column then forces string coercion, which is a
property of the format but means that the two encodings do not store these values in
the same representation (Section~\ref{sec:encoding}).

\paragraph{Internal validity.}
All MLT tiles were produced by one encoder version (maplibre-tiles 0.2.2) with default
options; other versions or settings, including future MLT encodings, may change absolute
savings. The archives were gzipped by different tools (GDAL and zlib), but the reported
gzip comparison recompresses both formats with identical level-6 settings, and the
stored lengths equal the recompressed ones. The equivalence of the paired encodings
rests on an audit that compares decoded content; its only differences are the two
format-related classes described in Section~\ref{sec:encoding}. String coercion can
matter for map styles whose expressions test property types, and
\ValidationUnchecked{} large tiles were checked only for addresses and sizes because the
MLT reference decoder exceeded the available memory.
The stream-level analysis relies on our own structural parser of the MLT~v1 framing.
After correcting the Boolean-RLE framing case, all parsed layer frames in all five
datasets and three variants complete without reported parse errors. The parser records
layer/column/stream boundaries and byte counts; it is used for structural statistics,
not as an independent semantic decoder. Structural results should therefore be regarded
as an implementation-dependent characterization of MLT~v1 and are published with the
analysis code. Tiles that PMTiles stores once for several addresses are counted at every
address for the headline size metrics, matching a client request for each address;
stream-structure statistics are collected once per stored payload. For this reason we do
not present the structural statistics as an exact algebraic decomposition of the
address-weighted granularity gain.

\paragraph{External validity.}
Five datasets cannot represent all vector maps, and each was clipped to a study area
that fits our processing resources. The PDOK and swisstopo study areas are small
(153 and 435 \VStd{} tiles), and at low zoom levels several study areas fit into one
tile, where densification cannot change anything. The datasets were chosen to differ in
production mode rather than sampled at random, and the results describe this corpus.
The systems that motivated the study (Section~\ref{sec:introduction}) were not
evaluated: no planetary data and no measurements on low-power servers are included.
MLT is evolving, and future versions may change both structural overhead and the
encoding efficiency of long streams.

\paragraph{Statistical conclusion validity.}
The analysis is descriptive. Tiles of one dataset are spatially dependent and nested
across zoom levels, so tile counts overstate the number of independent observations,
and we report no significance tests. Pooled bins are dominated by the largest dataset;
for this reason the main claims rest on within-dataset comparisons at matched logical
zoom levels.

\section{Conclusion}
\label{sec:conclusion}

Spatial tile granularity determines how much content an encoder can exploit inside one
independently encoded unit, and the experiment shows that this choice materially changes
the relative size of column-oriented MLT and feature-oriented MVT. Moving from the
conventional \VStd{} unit to the \VDsixteen{} unit increased MLT's raw saving over MVT
in all five datasets, with raw granularity gains from 1.0 to 7.3 percentage points. The
after-gzip gain remained positive in four of five datasets.

The strongest cross-dataset signal is not feature or vertex count alone but fragmentation
of the MLT representation. The five datasets are ordered exactly by their \VStd{} mean
values per stream: shorter streams correspond to larger raw granularity gains. Coarser
granularity lengthens streams and reduces the fraction of MLT layer-frame bytes occupied
by non-payload structure in every dataset; aggregate encoded payload bytes per value also
decrease from \VStd{} to \VDsixteen{} in all five. These results are consistent with two
mechanisms: amortization of layer/column/stream structure and, depending on the data,
more efficient encoding of longer value sequences.

For sparse datasets, coarser granularity also reduces bytes per stored vertex for MVT,
MLT and gzip alike, so the effect is not unique to one format. The broader implication is
that vector-tile format comparisons are conditional on the tiling granularity used in the
benchmark. Spatial granularity, coordinate extent, encoded byte size and rendered pixel
size should be reported separately. Coarser units are directly applicable where retrieval
granularity is secondary, such as offline archives; overzoom and variable-depth pyramids
are promising production mechanisms to investigate, but they are not equivalent to the
controlled matched-logical-zoom aggregation studied here, and their compression/overfetch
trade-off remains to be quantified.

For the kind of map that motivated this work, sparse custom thematic layers displayed
from stored archives on constrained hardware or links, the results point in one
direction: the fixed spatial unit of the conventional pyramid, not only the encoding
format, limits how compact such maps can be. The sparse datasets had the largest raw
granularity gains and the largest reductions in archive size, so for these maps the
tiling scheme is the first place to look for further compression.

\section*{Data Availability}
\label{sec:data-availability}

The processing pipeline and normalization profiles are archived as software
version~0.1.1 at \url{https://doi.org/10.5281/zenodo.23459428}.
The accompanying dataset of per-tile, per-layer and per-stream statistics
underlying the tables and figures is available at
\url{https://doi.org/10.5281/zenodo.23452598}.
Tile-level measurements are distributed as Parquet tables with schema metadata
and SQL view definitions; the accompanying analysis queries them directly using
DuckDB. The other inventories, diagnostics and materialized summaries are also
provided as SQLite databases with Parquet/CSV exports.
The source datasets are publicly available
from their providers under their respective licences.


\appendix
\section{Tile sizes in production web maps}
\label{app:traffic}

To see how large the tiles of production web maps are, we recorded the network traffic
of seven web products that show maps: OpenStreetMap, Yandex Maps, Amap, Cian, Airbnb,
Booking.com and Aviasales. Each product was loaded five times in desktop Chrome with a
cold cache and a viewport of $2560\times1600$ CSS pixels at a device pixel ratio of 1.5;
each run opened the map, panned by one map width and one map height and zoomed in once,
waiting for the network to be idle after each step. All requests, including those of web
workers and iframes, were recorded through the DevTools protocol, and tile bodies were
classified by decoding them. On our own archives, the recorded tile sizes matched the
stored sizes exactly (21 of 21 tiles).

Table~\ref{tab:traffic} lists the compressed sizes of individual tiles. The vector tiles
of Amap, Airbnb, Booking.com and Yandex Maps, all in proprietary formats, had medians
between 0.1 and 3.8~KB, and a page view loaded hundreds of them. The only MVT tiles
observed, from Aviasales, were world-level tiles at zoom~3 with a mean of 94~KB.

\begin{table}[!htbp]
\centering
\caption{Compressed size of individual tiles in seven web products (KB, median of runs;
1~KB = 1000 bytes). Measured on 2 October 2026 from one desktop machine.}
\label{tab:traffic}
\footnotesize
\setlength{\tabcolsep}{2.5pt}
\begin{tabular}{@{}l l r r r l@{}}
\toprule
Product & Format & Tiles & Median & P90 & Zoom \\
\midrule
OpenStreetMap & PNG              & 354 & 38.1 & 47.0  & 10, 14 \\
Cian          & PNG              & 363 & 16.3 & 27.5  & 10, 12 \\
Yandex Maps   & PNG + own vector & 677 & 3.8  & 25.9  & 9--11 \\
Amap          & own vector       & 774 & 1.2  & 5.4   & 10 \\
Airbnb        & Google vector    & 269 & 0.1  & 12.6  & 10--13 \\
Booking.com   & Google vector    & 357 & 2.0  & 5.8   & 11--16 \\
Aviasales     & MVT              & 8   & 95.7 & 157.2 & 3 \\
\bottomrule
\end{tabular}

\smallskip
{\scriptsize Tiles: tiles per run; own vector: proprietary vector format.}
\end{table}

The measurement has clear limits. It used one machine, one network location (Turkey),
one day and first loads only; sites localize by IP address, and street-level zoom was
reached only on OpenStreetMap and Booking.com. The formats of most products are
proprietary, so their sizes indicate the granularity at which vector map data are
delivered, not the size of MVT or MLT tiles with the same content.

\bibliographystyle{abbrvnat}
\raggedright
\bibliography{references}

\end{document}